\documentclass[11pt]{article}
\usepackage{vietnam,epsfig}

\def\Journal#1#2#3#4{{#1} {\bf #2}, #3 (#4)}

\def\PLB{{\em Phys. Lett.}  B}
\def\PRL{\em Phys. Rev. Lett.}
\def\PRD{{\em Phys. Rev.} D}
\def\JCAP{\em J. of Cosm. \& Astrop. Phys.} 
\def\CQG{\em Clas. \& Quant.\ Grav.}
\def\JHEP{\em J. of High Energy Phys.}

\def\lsim{\hbox{ \raise.35ex\rlap{$<$}\lower.6ex\hbox{$\sim$}\ }}
\def\gsim{\hbox{ \raise.35ex\rlap{$>$}\lower.6ex\hbox{$\sim$}\ }}

\def\xrightarrow#1#2#3#4{\,\lower#1pt\hbox{$\stackrel{\stackrel{\displaystyle #2}%
{\hbox to #3cm{\rightarrowfill}}}{#4}$}\,}
        
\def\be{\begin{equation}}
\def\ee{\end{equation}}
\def\bea{\begin{eqnarray}}
\def\eea{\end{eqnarray}}

\newcommand{\Mpl}{M_{\rm Pl}}

\def\lsim{\hbox{ \raise.35ex\rlap{$<$}\lower.6ex\hbox{$\sim$}\ }}
\def\gsim{\hbox{ \raise.35ex\rlap{$>$}\lower.6ex\hbox{$\sim$}\ }}

\begin{document}
\vspace*{4cm}
\title{PRODUCTION OF TOPOLOGICAL DEFECTS AT THE END OF INFLATION}

\author{ M. SAKELLARIADOU }

\address{Department of Physics, King's College, University of
London,\\ Strand, London WC2R 2LS, United Kingdom\\
Mairi.Sakellariadou@kcl.ac.uk}

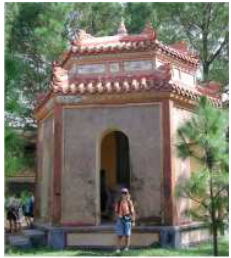
\maketitle \abstracts{Hybrid inflation within supersymmetric grand
unified theories, as well as inflation through brane collisions within
braneworld cosmological models, lead to the formation of
one-dimensional defects.  Observational data, mainly from the cosmic
microwave background temperature anisotropies but also from the
gravitational wave background, impose constraints on the free
parameters of the models. I review these inflationary models and
discuss the constraints from the currently available data.}

\section{Introduction}

The inflationary scenario is with no doubt extremely successful in
solving the shortcomings of the standard hot Big Bang model. Inflation
consists~\cite{infl1} of a phase of accelerated expansion which took
place at a very high energy scale. Inflation is deeply
rooted in the basic principles of general relativity and field theory,
while when the principles of quantum mechanics are taken into account,
it provides a successful explanation for the origin of the large
scale structures and the measured temperature anisotropies in the
Cosmic Microwave Background (CMB) spectrum.

However, despite its success inflation still remains a paradigm in
search of model. One should search for an inflationary model inspired
from some fundamental theory and subsequently test its predictions
against current data. This offers a beautiful and fruitful example of
the interplay between high energy physics and astrophysics/cosmology.
Inflation should also prove itself generic, meaning that the onset of
inflation should be independent of any particular initial conditions.
This issue, already addressed in the past~\cite{piran,calzettamairi},
has been recently re-investigated~\cite{GT,GNS}.

Hybrid inflation is based on Einstein's gravity but driven by false
vacuum. The inflaton field rolls down its potential while another
scalar field is trapped in an unstable false vacuum. Once the inflaton
field becomes much smaller than some critical value, a phase
transition to the true vacuum takes place signaling the end of
inflation. Such phase transition may leave behind topological defects
as false vacuum remnants.

In this lecture, I will first discuss hybrid inflation within
Supersymmetric Grand Unified Theories (SUSY GUTs) and then within
braneworld cosmological models. In both cases one-dimensional
topological defects are generically produced at the end of
inflation. I will discuss the predictions of these models, in
particular regarding the CMB temperature anisotropies spectrum but 
also with respect to the gravitational wave background, which
consequently induce constraints on the parameters space of the models.

\section{Inflation within supersymmetric grand unified theories}

Theoretically motivated inflationary models can be built in the
context of Supersymmetry (SUSY) or Supergravity (SUGRA).
$N=1$ SUSY models contain complex scalar fields which often
have flat directions in their potential, thus offering natural
candidates for inflationary models. In this framework, hybrid
inflation driven by F-terms or D-terms is the standard
inflationary model, leading~\cite{JRS} generically to cosmic
string formation at the end of inflation.  F-term inflation is
potentially plagued with the $\eta$-problem, while D-term inflation
avoids it.

\subsection{F-term inflation}
F-term inflation can be naturally accommodated in the framework of
GUTs when a GUT gauge group, G$_{\rm GUT}$, is broken down to the
Standard Model (SM) gauge group, G$_{\rm SM}$, at an energy scale
$M_{\rm GUT}$ according to the scheme
\begin{equation}
{\rm G}_{\rm GUT} \stackrel{M_{\rm GUT}}{\hbox to 0.8cm
{\rightarrowfill}} {\rm H}_1 \xrightarrow{9}{M_{\rm
infl}}{1}{\Phi_+\Phi_-} {\rm H}_2 {\longrightarrow} {\rm G}_{\rm SM}~,
\end{equation}
where $\Phi_+, \Phi_-$ is a pair of GUT Higgs superfields in non-trivial
complex conjugate representations, which lower the rank of the group
by one unit when acquiring non-zero vacuum expectation value. The
inflationary phase takes place at the beginning of the symmetry
breaking ${\rm H}_1\stackrel{M_{\rm infl}}{\longrightarrow} {\rm
H}_2$.
The gauge symmetry is spontaneously broken by adding F-terms to the
superpotential. The Higgs mechanism leads generically~\cite{JRS} to
Abrikosov-Nielsen-Olesen strings, called F-term strings.

F-term inflation is based on the globally supersymmetric
renormalisable superpotential
\begin{equation}\label{superpot}
W_{\rm infl}^{\rm F}=\kappa  S(\Phi_+\Phi_- - M^2)~,
\end{equation}
where $S$ is a GUT gauge singlet left handed superfield and $\kappa$, $M$
 are two constants ($M$ has dimensions of mass) which can be taken
 positive with field redefinition.  The scalar potential as a function
 of the scalar complex component of the respective chiral superfields
 $\Phi_\pm, S$ reads
\begin{equation}
\label{scalpot1}
V(\phi_+,\phi_-, S)= |F_{\Phi_+}|^2+|F_{\Phi_-}|^2+|F_ S|^2
+\frac{1}{2}\sum_a g_a^2 D_a^2~.
\end{equation}
The F-term is such that $F_{\Phi_i} \equiv |\partial W/\partial
\Phi_i|_{\theta=0}$, where we take the scalar component of the
superfields once we differentiate with respect to $\Phi_i=\Phi_\pm,
 S$. The D-terms are
$D_a=\bar{\phi}_i\,{(T_a)^i}_j\,\phi^j +\xi_a$,
with $a$ the label of the gauge group generators $T_a$, $g_a$ the
gauge coupling, and $\xi_a$ the Fayet-Iliopoulos term. By
definition, in the F-term inflation the real constant $\xi_a$ is zero;
it can only be non-zero if $T_a$ generates an extra U(1) group.  In the
context of F-term hybrid inflation the F-terms give rise to the
inflationary potential energy density while the D-terms are flat
along the inflationary trajectory, thus one may neglect them during
inflation.

The potential has one valley of local minima, $V=\kappa^2 M^4$, for
$S> M $ with $\phi_+ = \phi_-=0$, and one global supersymmetric
minimum, $V=0$, at $S=0$ and $\phi_+ = \phi_- = M$. Imposing initially
$ S \gg M$, the fields quickly settle down the valley of local
minima.  Since in the slow roll inflationary valley the ground state
of the scalar potential is non-zero, supersymmetry is broken.  In the tree
level, along the inflationary valley the potential is constant,
therefore perfectly flat. A slope along the potential can be generated
by including the one-loop radiative corrections. Thus, the scalar
potential gets a little tilt which helps the inflaton field $S$ to
slowly roll down the valley of minima. The one-loop radiative
corrections to the scalar potential along the inflationary valley
lead to the effective potential~\cite{rs1}
\begin{eqnarray}
\label{VexactF}
V_{\rm eff}^{\rm F}(|S|)&=&\kappa^2M^4\biggl\{1+\frac{\kappa^2
\cal{N}}{32\pi^2}\biggl[2\ln\frac{|S|^2\kappa^2}{\Lambda^2}
+(z+1)^2
\ln(1+z^{-1})\nonumber\\
&&~~~~~~~~~+(z-1)^2\ln(1-z^{-1})
\biggr]\biggr\} ~~; ~~z=\frac{|S|^2}{M^2}~,
\end{eqnarray}
$\Lambda$ is a renormalisation scale
and $\cal{N}$ stands for the dimensionality
of the representation to which the complex scalar components $\phi_+,
\phi_-$ of the chiral superfields $\Phi_+, \Phi_-$ belong.

\subsection{D-term inflation}
D-term inflation is one of the most interesting and versatile models
of inflation. It is possible to implement naturally D-term inflation
within high energy physics, as for example SUSY GUTs, SUGRA, or string
theories.  In addition, D-term inflation avoids the {\sl
Hubble-induced mass} problem. Here, the gauge symmetry is
spontaneously broken by introducing Fayet-Iliopoulos (FI) D-terms. In
standard D-term inflation, the constant FI term gets compensated by a
single complex scalar field at the end of the inflationary era, which
implies that standard D-term inflation ends with the formation of
cosmic strings, called D-strings.  A supersymmetric description of the
standard D-term inflation is insufficient; the inflaton field reaches
values of the order of the Planck mass, or above it, even if one
concentrates only around the last 60 e-folds of inflation; the correct
analysis is therefore in the context of
supergravity.

Standard D-term inflation
requires a scheme
\begin{equation}
{\rm G}_{\rm GUT}\times {\rm U}(1) \stackrel{M_{\rm GUT}}{\hbox to
0.8cm{\rightarrowfill}} {\rm H} \times {\rm U}(1) \xrightarrow{9}{M_{\rm
nfl}}{1}{\Phi_+\Phi_-} {\rm H} \rightarrow {\rm G}_{\rm SM}~.
\end{equation}
It is based on the superpotential
\begin{equation}\label{superpoteninflaD}
W=\lambda S\Phi_+\Phi_-~,
\end{equation}
where $S, \Phi_+, \Phi_-$ are three chiral superfields and $\lambda$
is the superpotential coupling. It assumes an invariance under an
Abelian gauge group $U(1)_\xi$, under which the superfields $S,
\Phi_+, \Phi_-$ have charges $0$, $+1$ and $-1$, respectively. It
also assumes the existence of a constant FI term $\xi$.  
In the \emph{standard} supergravity formulation 
the Lagrangian depends on the K\"ahler potential
$K(\Phi_i,\bar{\Phi}_i)$ and the superpotential $W(\Phi_i)$ only
through the combination given in
\begin{equation}
\label{kwcomb}
G(\Phi_i,\bar{\Phi}_i)= \frac{K(\Phi_i,\bar{\Phi}_i)}{\Mpl^2} +\ln
\frac{|W(\Phi_i)|^2}{\Mpl^6}~.
\end{equation}
However, this \emph{standard} supergravity formulation is
inappropriate to describe D-term inflation~\cite{toine1}. In D-term
inflation the superpotential vanishes at the unstable de Sitter vacuum
(anywhere else the superpotential is non-zero). Thus, \emph{standard}
supergravity is inappropriate, since the theory is ill-defined at
$W=0$. In conclusion, D-term inflation must be described with a
non-singular formulation of supergravity when the superpotential
vanishes.

Various formulations of effective supergravity can be constructed from
the superconformal field theory. One must first build a Lagrangian
with full superconformal theory, and then the gauge symmetries that
are absent in Poincar\'e supergravity must be gauge fixed. In this
way, one can construct a non-singular theory at $W=0$, where the
action depends on all three functions: the K\"ahler potential
$K(\Phi_i,\bar{\Phi}_i)$, the superpotential $W(\Phi_i)$ and the
kinetic function $f_{ab}(\Phi_i)$ for the vector multiplets.
To construct a formulation of supergravity with constant FI terms from
superconformal theory, one finds~\cite{toine1} that under U(1) gauge
transformations in the directions in which there are constant FI terms
$\xi_\alpha$, the superpotential $W$ must transform as 
$\delta_\alpha W=\eta_{\alpha i}\partial^i W = -i
(g\xi_\alpha/\Mpl^2)W$;
one cannot keep any longer the same charge assignments as in
standard supergravity.

D-term inflationary models can be built with different choices of the
K\"ahler geometry.  Let us first consider D-term inflation within minimal
supergravity. It is based on
\begin{equation}\label{Kmin}
K_{\rm min}=\sum_i |\Phi_i|^2=|\Phi_-|^2+|\Phi_+|^2+|S|^2~,
\end{equation}
with $f_{ab}(\Phi_i)=\delta_{ab}$. 
The tree level scalar potential is~\cite{toine1}
\begin{eqnarray}\label{DpotenSUGRAtotbis}
V_{\rm min}=&&
\lambda^2\exp\left({\frac{|\phi_-|^2+|\phi_+|^2+|S|^2}{M^2_{\rm
Pl}}}\right) 
\Biggl[|\phi_+\phi_-|^2\left(1+\frac{|S|^4}{M^4_{\rm
Pl}}\right)+|\phi_+S|^2 \left(1+\frac{|\phi_-|^4}{M^4_{\rm
Pl}}\right)\nonumber\nonumber\\
 && 
+|\phi_-S|^2 \left(1+\frac{|\phi_+|^4}{M^4_{\rm
Pl}}\right) +3\frac{|\phi_-\phi_+S|^2}{M^2_{\rm Pl}}\Biggr]\
+ \frac{g^2}{2}\left(q_+|\phi_+|^2+q_-|\phi_-|^2+\xi\right)^2~,
\end{eqnarray}
with
$q_\pm = \pm 1-\xi/(2\Mpl^2).$
The potential has two minima: One global minimum at zero and one local
minimum equal to $V_0=(g^2/2)\xi^2$.  For arbitrary large $S$ the tree
level value of the potential remains constant and equal to
$V_0$; the $S$ plays the r\^ole of the inflaton
field. Assuming chaotic initial conditions $|S|\gg S_{\rm s}$,
inflation begins. Along the inflationary trajectory the D-term, which is the
dominant one, splits the masses in the $\Phi_\pm$ superfields,
leading to the one-loop effective potential for the inflaton field.
Considering also the one-loop radiative corrections~\cite{rs1,prl2005}
\begin{equation}\label{scalarpeff}
V^{\rm eff}_{\rm min}(|S|)=\frac{g^2\xi^2}{2}\left\{
1+\frac{g^2}{16\pi^2}\left[2\ln\left(z\frac{g^2\xi}{\Lambda^2}\right)+f_V(z)
\right] \right\} ~,
\end{equation}
where
\begin{equation}
f_V(z) = (z+1)^2\ln\left( 1+\frac{1}{z}\right) + (z-1)^2\ln\left(
1-\frac{1}{z}\right)~~,
~~z\equiv \frac{\lambda^2}{g^2\xi} |S|^2
\exp\left(\frac{|S|^2}{M_{\rm Pl}^2}\right)~.
\end{equation}

As a second example, consider D-term inflation based on K\"ahler
geometry with shift symmetry,
\begin{equation}\label{K3}
K_{\rm shift}=\frac{1}{2} (S+\bar{S})^2+|\phi_+|^2+|\phi_-|^2~,
\end{equation}
and minimal structure for the kinetic function. 
The scalar potential reads~\cite{rs3}
\begin{eqnarray}
V_{\rm shift}&\simeq&
~\frac{g^2}{2}\left(|\phi_+|^2-|\phi_-|^2+\xi\right)^2 \nonumber\\
&&+\lambda^2\exp\left({\frac{|\phi_-|^2+|\phi_+|^2}{M^2_{\rm
Pl}}}\right)\exp\left[{\frac{(S+\bar{S})^2}{2M^2_{\rm Pl}}}\right]
\nonumber\\ & & ~~~\times
\Biggl[|\phi_+\phi_-|^2\left(1+\frac{S^2+\bar{S}^2}{M^2_{\rm
Pl}}+\frac{|S|^2|S+\bar{S}|^2}{M^4_{\rm Pl}}\right)+|\phi_+S|^2
\left(1+\frac{|\phi_-|^4}{M^4_{\rm Pl}}\right) \nonumber\\ &&
+|\phi_-S|^2 \left(1+\frac{|\phi_+|^4}{M^4_{\rm Pl}}\right)
+3\frac{|\phi_-\phi_+S|^2}{M^2_{\rm Pl}}\Biggr] ~.
\end{eqnarray}
As in D-term inflation within minimal SUGRA, the potential has a
global minimum at zero for $\langle\Phi_+\rangle=0$ and
$\langle\Phi_-\rangle=\sqrt{\xi}$ and a local minimum equal to
$V_0=(g^2/2)\xi^2$ for $\langle S\rangle\gg S_{\rm c}$ and
$\langle\Phi_\pm\rangle=0$.  The exponential factor $e^{|S|^2}$ in
minimal SUGRA has been replaced by $e^{(S+\bar{S})^2/2}$.  Writing
$S=\eta+i\phi_0$ one gets $e^{(S+\bar{S})^2/2}=e^{\eta^2}$.  If
$\eta$ plays the r\^ole of the inflaton field, we obtain the same
potential as for minimal D-term inflation. If instead $\phi_0$ is the
inflaton field, the inflationary potential is identical to that of the
usual D-term inflation within global SUSY~\cite{rs1}.  The latter case
is better adapted with the choice $K_{\rm shift}$, since in this case the
exponential term is constant during inflation, thus it cannot spoil
the slow roll conditions.

As a last example, consider a K\"ahler potential with
non-renormalisable terms:
\begin{equation}
K_{\rm
non-renorm}=|S|^2+|\Phi_+|^2+|\Phi_-|^2+f_+\bigg(\frac{|S|^2}{M_{\rm
Pl}^2}\bigg)|\Phi_+|^2+f_-\bigg(\frac{|S|^2}{M_{\rm Pl}^2}\bigg)
|\Phi_-|^2+b\frac{|S|^4}{\Mpl^2}~,
\label{gen}
\end{equation}
where $f_\pm$ are arbitrary functions of $(|S|^2/M_{\rm Pl}^2)$ and
the superpotential is given in Eq.~(\ref{superpoteninflaD}).  The
effective potential reads
\begin{equation}
V^{\rm eff}_{\rm non-renorm}(|S|)=\frac{g^2\xi^2}{2}\left\{
1+\frac{g^2}{16\pi^2}\left[ 2\ln \left(
z\frac{g^2\xi}{\Lambda^2}\right)+f_V(z) \right] \right\} ~,
\end{equation}
where \begin{equation}\label{deffV}
f_V(z) = (z+1)^2\ln\left( 1+\frac{1}{z}\right) + (z-1)^2\ln\left(
1-\frac{1}{z}\right)~
\end{equation}
\begin{equation}
\mbox{with} ~~~~z\equiv
\frac{\lambda^2|S|^2}{g^2\xi}\exp\bigg(\frac{|S|^2}{\Mpl^2}
+b\frac{|S|^4}{\Mpl^4}\bigg) \frac{1}{(1+f_+)(1+f_-)}~.\\
\end{equation}

\section{Inflation within braneworld cosmologies}
In the context of braneworld cosmology, brane
inflation occurs in a similar way as hybrid inflation within
supergravity, leading to string-like objects.  In string
theories, D-brane $\bar{\rm D}$-anti-brane annihilation leads
generically to the production of lower dimensional D-branes, with D3-
and D1-branes (D-strings) being predominant~\cite{rmm}.

To sketch brane inflation (for example see~\cite{tye}), consider a
D$p$-${\bar{\rm D}}p$ system in the context of IIB string theory. Six
of the spatial dimensions are compactified on a torus; all branes move
relatively to each other in some directions. A simple and
well-motivated inflationary model is brane inflation where the
inflaton is simply the position of a D$p$-brane moving in the bulk.
As two branes approach, the open string modes between the branes
develop a tachyon, indicating an instability.  The relative
D$p$-${\bar{\rm D}}p$-brane position is the inflaton field and the
inflaton potential comes from their tensions and interactions.  Brane
inflation ends by a phase transition mediated by open string
tachyons. The annihilation of the branes releases the brane tension
energy that heats up the universe so that the hot big ban epoch can
take place. Since the tachyonic vacuum has a non-trivial $\pi_1$
homotopy group, there exist stable tachyonic string solutions with
$(p-2)$ co-dimensions. These daughter branes have all dimensions
compact; a four-dimensionnal observer perceives them as
one-dimensional objects, the D-strings.  Zero-dimensional defects
(monopoles) and two-dimensional ones (domain walls), which are
cosmologically undesirable, are not produced during brane
intersections.  Fundamental strings (F-strings) and D-strings that
survive the cosmological evolution become cosmic superstrings.  In
these models, the large compact dimensions and the large warp factors
allow cosmic superstring tensions to be in the range between
$10^{-13}< G\mu < 10^{-6}$, depending on the model. These cosmic
suprestrings are stable, or at least their lifetime is comparable to
the age of the universe, so they can survive to form a cosmic
superstring network.

Cosmic superstrings share a number of properties with the cosmic
strings, but there are also differences which may lead to distinctive
observational signatures.  When F- and D-strings meet they can form a
three-string junction, with a composite FD-string.  IIB string theory
allows the existence of bound $(p,q)$ states of $p$ F-strings and $q$
D-strings, where $p$ and $q$ are coprime.  String intersections lead
to intercommutation and loop production. For cosmic strings the
probability of intercommutation ${\cal P}$ is equal to 1, whereas this
is not the case for F- and D-strings. Clearly, D-strings can miss each
other in the compact dimension, leading to a smaller ${\cal P}$, while
for F-strings the scattering has to be calculated quantum mechanically
since these are quantum mechanical objects. The evolution of cosmic
superstring networks has been studied numerically. All
studies~\cite{ms,ep,tww} conclude that the network will reach a {\sl
scaling} regime.

Cosmic superstrings interact with the standard model particles only
via gravity, implying that their detection involves gravitational
interactions. Since the particular brane inflationary scenario remains
unknown, the tensions of superstrings are only loosely constrained.

\section{Compatibility between predictions and data}

\subsection{CMB temperature anisotropies}
Cosmic strings and string-like objects are expected to be generically
formed within a large class of models (SUSY, SUGRA and string
theories), thus one should consider {\sl mixed perturbation models}
where the dominant r\^ole is played by the inflaton field and cosmic
strings have a small, but not negligible, contribution.  Restricting
ourselves to the angular power spectrum we remain in the linear
regime, where
\begin{equation}
C_\ell =   \alpha     C^{\scriptscriptstyle{\rm I}}_\ell
         + (1-\alpha) C^{\scriptscriptstyle{\rm S}}_\ell~;
\label{cl}
\end{equation}
$C^{\scriptscriptstyle{\rm I}}_\ell$ and $C^{\scriptscriptstyle
{\rm S}}_\ell$ denote the (COBE normalized) Legendre coefficients due
to adiabatic inflaton fluctuations and those stemming from the cosmic
strings network, respectively. The coefficient $\alpha$ in
Eq.~(\ref{cl}) is a free parameter giving the relative amplitude for
the two contributions.  Comparing the $C_\ell$, given by
Eq.~(\ref{cl}), with data obtained from the most recent CMB
measurements, one can impose constraints on
the parameters space of the models. The upper limit
imposed~\cite{bprs,fraisse,pog2} on the cosmic string contribution to
the CMB data depends on the numerical simulation employed in order to
calculate the cosmic string power spectrum. In what follows I will not
allow cosmic strings to contribute more than $9\%$ to the CMB
temperature anisotropies.

\subsubsection{F-term inflation}
Considering only large angular scales one can get the contributions
to the CMB temperature anisotropies analytically.  The quadrupole
anisotropy has one contribution coming from the inflaton field,
calculated using Eq.~(\ref{VexactF}), and one contribution coming from
the cosmic strings network.  Fixing the number of e-foldings to 60, 
the inflaton and cosmic strings contribution to the CMB depend on the
superpotential coupling $\kappa$, or equivalently on the symmetry
breaking scale $M$ associated with the inflaton mass scale, which
coincides with the string mass scale. The total quadrupole anisotropy
has to be normalised to the COBE data.  The cosmic strings
contribution is consistent with the CMB measurements
provided~\cite{rs1}
\begin{equation}
M\lsim 2\times 10^{15} {\rm GeV} ~~\Leftrightarrow ~~\kappa \lsim
7\times10^{-7}~.
\end{equation}

The superpotential coupling $\kappa$ is also subject to the gravitino
constraint which imposes an upper limit to the reheating temperature,
to avoid gravitino overproduction. Within the framework of SUSY GUTs
and assuming a see-saw mechanism to give rise to massive neutrinos,
the inflaton field decays during reheating into pairs of right-handed
neutrinos.  This constraint on the reheating temperature can be
converted to a constraint on the parameter $\kappa$. The gravitino
constraint on $\kappa$ reads~\cite{rs1} $\kappa \lsim 8\times
10^{-3}$, which is a weaker constraint.

The tuning of the free parameter $\kappa$ can be softened if one
allows for the curvaton mechanism.  The curvaton is a scalar field
that is sub-dominant during the inflationary era as well as at the
beginning of the radiation dominated era which follows the
inflationary phase. There is no correlation between the primordial
fluctuations of the inflaton and curvaton fields. Clearly, within
supersymmetric theories such scalar fields are expected to exist. In
addition, embedded strings, if they accompany the formation of cosmic
strings, they may offer a natural curvaton candidate, provided the
decay product of embedded strings gives rise to a scalar field before
the onset of inflation.  Considering the curvaton scenario, the
coupling $\kappa$ is only constrained by the gravitino limit. More
precisely, assuming the existence of a curvaton field there is an
additional contribution to the temperature anisotropies. The WMAP CMB
measurements impose~\cite{rs1} the following limit on the initial
value of the curvaton field
\begin{equation}
{\cal\psi}_{\rm init} \lsim 5\times 10^{13}\,\left( 
\frac{\kappa}{10^{-2}}\right){\rm GeV}~~\mbox{for}~~
 \kappa\in [10^{-6},~1].
\end{equation}

\subsubsection{D-term inflation}
D-term inflation leads to cosmic string formation at the end of the
inflationary era. In the case of minimal SUGRA, consistency between
CMB measurements and theoretical predictions impose~\cite{rs1,prl2005}
that $g\lsim 2\times 10^{-2}~ ~ \mbox{and}~ ~ \lambda\lsim 3\times
10^{-5}$, which can be expressed as a single constraint on the
Fayet-Iliopoulos term $\xi$, namely $\sqrt\xi \lsim
2\times 10^{15}~{\rm GeV}.$ Note that for minimal D-term inflation one
can neglect the corrections introduced by the superconformal origin of
supergravity.  Within minimal SUGRA the couplings and masses
must be fine tuned to achieve compatibility between measurements on
the CMB anisotropies and theoretical predictions.  The fine tuning on
the couplings can be softened if one invokes the curvaton mechanism
and provided the initial value of the curvaton field 
is~\cite{prl2005}
\begin{equation}
\psi_{\rm init}\lsim 3\times 10^{14}\left(\frac{g}{10^{-2}}\right){\rm
GeV}~~\mbox{for}~~\lambda\in[10^{-1},10^{-4}]~.
\end{equation}

The constraints on the couplings remain qualitatively
valid in non-minimal SUGRA theories, with the superpotential $W$ given
in Eq.~(\ref{superpoteninflaD}) and a non-minimal K\"ahler potential.
Let us first consider D-term inflation
based on K\"ahler geometry with shift symmetry.
The cosmic string contribution to the CMB anisotropies is dominant, in
contradiction with the CMB measurements, unless the superpotential
coupling is~\cite{rs3}
$\lambda\lsim 3\times 10^{-5}.$
Also in the case of D-term inflation based on a K\"ahler potential with
non-renormalisable terms,
the contribution of cosmic strings dominates
if the superpotential coupling $\lambda$ is close to unity.  The
constraints on $\lambda$ read~\cite{rs3}
\begin{equation}\label{contraintelambdanonmin}
(0.1-5)\times 10^{-8} 
 \leq \lambda \leq
(2-5)\times 10^{-5} 
\mbox{~or, equivalently~}
\sqrt{\xi}\leq 2\times 10^{15}
\;\mathrm{GeV}~,
\end{equation}
implying $ G\mu \leq 8.4\times 10^{-7}$. In conclusion, higher order
K\"ahler potentials do {\bf not} suppress cosmic string contribution,
as it was incorrectly claimed in the literature.

\subsubsection{Brane inflation}
The CMB temperature anisotropies originate from the
amplification of quantum fluctuations during inflation, as well as
from the cosmic superstring network. Provide the scaling regime of the
superstring netwotk is the unique source of the density perturbation,
the COBE data imply $G\mu \simeq 10^{-6}$. The latest WMAP data allow
at most a 9$\%$ contribution from strings, which imply a bound
$G\mu\lsim 7\times 10^{-7}$. Thus, the cosmic superstrings produced
towards the end of inflation in the context of braneworld cosmological
models is in agreement with the present CMB data.

Strings/superstrings moving with velocity ${\bf v}$ across the sky
lead to a shift $\Delta T/T\simeq 8\pi G\mu v\gamma$ in the CMB
temperature. Current CMB data may probe $G\mu\simeq 10^{-10}$; this
limit may be possible to go down to $G\mu\simeq 10^{-13}$.

\subsection{Gravitational wave background}

Oscillating cosmic string loops emit~\cite{vil} Gravitational Waves
(GW). Long strings are not staight but they have a superimposed wiggly
small-scale structure due to string intercommutations, thus they also
emit~\cite{ms-gw} GW. Cosmic superstrings can as well
generate~\cite{dv} a stochastic GW background. Therefore, provided the
emission of gravity waves is the efficient
mechanism~\cite{gwandstrings} for the decay of string loops, cosmic
strings/superstrings could provide a source for the stochastic GW
spectrum in the low-frequency band.  The stochastic GW spectrum has an
almost flat region in the frequency range $10^{-8}-10^{10}$ Hz. Within
this window, both ADVANCED LIGO/VIRGO (sensitive at a frequency $f\sim
10^2$ Hz) and LISA (sensitive at $f\sim 10^{-2}$ Hz) interferometers
may have a chance of detectability.

Strongly focused beams of relatively high-frequency GW are emitted by
cusps and kinks in oscillating strings/superstrings. The distinctive
waveform of the emitted bursts of GW may be the most sensitive test of
strings/superstrings. ADVANCED LIGO/VIRGO may detect bursts of GW for
values of $G\mu$ as low as $10^{-13}$, and LISA for values down to
$G\mu\geq 10^{-15}$. At this point, I would like to remind to the
reader that there is still a number of theoretical uncertainties for
the evolution of a string/superstring network.

Recently, they have been imposed limits~\cite{pulsar} on an isotropic
gravitational wave background using pulsar timing observations, which
offer a chance of studying low-frequency (in the range between
$10^{-9}-10^{-7}$ Hz) gravitational waves.  The imposed limit on the
energy density of the background per unit logarithmic frequency
interval reads $\Omega_{\rm gw}^{\rm cs}(1/8{\rm yr})h^2\leq 1.9\times
10^{-8}$ (where $h$ stands for the dimensionless amplitude in GW
bursts).  If the source of this isotropic GW background is a cosmic
string/superstring network, then it leads to an upper bound on the
dimensionless tension of a cosmic string background. Under reasonable
assumptions for the string network the upper bound on the string
tension reads~\cite{pulsar} $G\mu\leq 1.5\times 10^{-8}$. This is a
strongest limit than the one imposed from the CMB temperature
anisotropies. Clearly, to achieve compatibility with this constraint,
F- and D-term inflation become even more fine tuned, unless one
invokes the curvaton mechanism.  This limit does not affect cosmic
superstrings. However, it has been argued~\cite{pulsar} that with the
full Parkes Pulsar Timing Array (PPTA) project the upper bound will
become $G\mu\leq 5\times 10^{-12}$, which is directly relevant for
cosmic superstrings. In conclusion, the full PPTA will either detect
gravity waves from strings or they will rule out a number of models.

\section{Conclusions}
Cosmic strings and string-like objects are generically formed at the
end of inflation in the framework of SUSY GUTs as well as in the
context of branewold models.  These objects contribute to the CMB
temperature anisotropies implying constraints on the parameters of the
models. More precisely, by allowing a small but non-negligible
contribution of strings to the angular power spectrum of CMB
anisotropies, we constrain the couplings of the inflationary models,
or equivalently the dimensionless string tension. These models remain
compatible with the most current CMB measurements, even when we
calculate~\cite{ns} the spectral index. Namely, the inclusion of a
sub-dominant string contribution to the large scale power spectrum
amplitude of the CMB increases the preferred value for the spectral
index.  In addition, cosmic strings/superstrings can contribute to the
stochastic gravitational background, thus limits can be imposed to the
dimensionless string tension; these limits are indeed strongest.

F-/D-term inflation within SUSY GUTs or branewold inflation leading to
string-like defects are inflationary models which are definitely {\bf
not} ruled out. Stable strings/superstrings are indeed generically
produced in the simplest versions of these models, which then become
severly constrained from currently available data. However, one can
include additional ingredients so that the one-dimensional defects
formed at the end of inflation become unstable~\cite{toine1,unstable}.
This may be indeed the most realistic approach to model building.

\section*{Acknowledgments}
It is a pleasure to thank the organisers of the Conference
``Challenges in Particle Astrophysics'' $6^{\rm th}$ {\sl Rencontres
du Vietnam} for inviting me to present this work. I acknowledge
financial support from the Royal Society conference grants scheme
(Conference Grant 2006/R1).


\section*{References}
\begin{thebibliography}{99}
\bibitem{infl1}
A.\ Guth, \Journal{\PRD}{23}{347}{1981}.

\bibitem{piran}
T.\ Piran, \Journal{PLB}{181}{238}{1986}; D.\ S.\ Goldwirth,
\Journal{\PRD}{43}{3204}{1991}.

\bibitem{calzettamairi}
E.\ Calzetta and M.\ Sakellariadou, \Journal{\PRD}{45}{2802}{1992};
E.\ Calzetta and M.\ Sakellariadou, \Journal{\PRD}{47}{3184}{1993}.

\bibitem{GT} G.\ W.\ Gibbons and N.\ Turok, {\sl The measure problem
in cosmology}, [hep-th/0609095].

\bibitem{GNS} C.\ Germani, W.\ Nelson and M.\ Sakellariadou,
{\sl On the onset of inflation in loop quantum cosmology}, [gr-qc/0701172].

\bibitem{JRS}
R.\ Jeannerot, J.\ Rocher, M.\ Sakellariadou,
\Journal{\PRD}{68}{103514}{2003}.

\bibitem{rs1}
J.\ Rocher and M.\ Sakellariadou, \Journal{\JCAP}{0503}{004}{2005}.

\bibitem{toine1} P.~ Binetruy, G.~ Dvali, R.~ Kallosh and A.~ Van
Proeyen, \Journal{\CQG}{21}{3137}{2004}.

\bibitem{prl2005} J.\ Rocher, M.\ Sakellariadou,
\Journal{\PRL}{94}{011303}{2005}.

\bibitem{rs3}
J.\ Rocher and M.\ Sakellariadou, \Journal{\JCAP}{0611}{001}{2006}.

\bibitem{rmm} R.\ Durrer, M.\ Kunz and M.\ Sakellariadou, 
\Journal{\PLB}{614}{12}{2005}.

\bibitem{tye}
S.-H.\ Tye, \emph{Brane inflation: string theory viewed from the cosmos},
\texttt{[hep-th/0610221]}.

\bibitem{ms} M.\ Sakellariadou, \Journal{\JCAP}{0504}{003}{2005}.

\bibitem{ep}
E.\ Copeland and P.\ Saffin, \Journal{\JHEP}{0511}{023}{2005}.

\bibitem{tww} S.-H.\ H.\ Tye, I.\ Wasserman and M.\ Wyman,
\Journal{\PRD}{71}{103508}{2005};
Erratum-ibid. \Journal{\PRD}{71}{129906}{2005}.
 
\bibitem{bprs} F.\ R.\ Bouchet, P.\ Peter, A.\ Riazuelo and M.\
Sakellariadou, \Journal{\PRD}{65}{021301}{2002}.

\bibitem{fraisse} A.\ A.\ Fraisse, \emph{Limits on SUSY GUTs and defects
formation in hybrid inflationary models with three-year WMAP
observations}, \texttt{[astro-ph/0603589]}.

\bibitem{pog2} L.\ Pogosian, S.\ H.\ Tye, I.\ Wasserman and M.\ Wyman,
Erratum-ibid. \Journal{\PRD}{73}{089904}{2006}.

\bibitem{vil}
A.~ Vilenkin, \Journal{\PLB}{107}{47}{1981}

\bibitem{ms-gw}
M.~ Sakellariadou, \Journal{\PRD}{42}{354}{1990}.

\bibitem{dv}
T.~ Damour and A.~ Vilenkin, \Journal{\PRD}{71}{063510}{2005}.

\bibitem{gwandstrings} G.\ R.\ Vincent, M.\ Hindmarsh and M.\
Sakellariadou, \Journal{\PRD}{56}{637}{1997}; G.\ R.\ Vincent, N.\ D.\
Antunes and M.\ Hindmarsh, \Journal{\PRL}{80}{2277}{1998};
C.\ Ringeval, M.\ Sakellariadou and F.\ R.\ Bouchet,
\emph{Cosmological evolution of cosmic string loops},
\texttt{[astro-ph/0511646]}.

\bibitem{pulsar} F.~ A.~ Jenet, {\sl et al.}, \emph{Upper bounds on
the low-frequency stochastic gravitational wave background from pulsar
timing observations: current limits and future prospects},
\texttt{[astro-ph/0609013]}.

\bibitem{ns} R.\ A.\ Battye, B.\ Garbrecht and A.\ Moss,
\Journal{\JCAP}{0609}{007}{2006}.

\bibitem{unstable} J.\ Urrestilla, A.\ Ach{\'u}carro and A.\ C.\
Davis, \Journal{\PRL}{92}{251302}{2004}.


\end{thebibliography}
 \end{document}